\documentclass[twocolumn, a4paper,10pt]{article}
\usepackage[top=2.5cm, bottom=2.5cm, left=2.0cm, right=2.0cm,
columnsep=0.8cm]{geometry}
\usepackage{lipsum}
\usepackage{enumitem}
\usepackage[hidelinks]{hyperref}
\usepackage{xurl}
\usepackage{amsmath,amsfonts,amssymb}
\usepackage{gensymb}
\usepackage[bottom]{footmisc}
\usepackage{boxedminipage}
\usepackage{nopageno}
\usepackage{graphicx}
\usepackage{natbib}
\usepackage[font=it]{caption}
\usepackage[usenames,dvipsnames]{xcolor}
\usepackage{listings}
\usepackage{caption}
\usepackage{subcaption}
\setlist{itemsep=-0.1cm,topsep=0.1cm,labelsep=0.3cm}
\setenumerate{leftmargin=*}
\makeatletter
\renewcommand\title[1]{\gdef\@title{\fontsize{12pt}{2pt}\bfseries{#1}}}
\makeatletter
\renewcommand\section{\@startsection{section}{1}{\z@}{3pt}{3pt}{\normalfont\large\bfseries}}
\makeatletter
\renewcommand\subsection{\@startsection{subsection}{1}{\z@}{\z@}{\z@}{\normalfont\normalsize\bfseries}}
\makeatletter
\renewcommand\subsection{\@startsection{subsection}{1}{\z@}{\z@}{0.1pt}{\normalfont\normalsize\bfseries}}

\title{%
Value of Information Analysis for rationalising\\
\vspace{4pt}
information gathering in building energy analysis}
\author{
Max Langtry$^1$, Chaoqun Zhuang$^{1,2}$, Rebecca Ward$^{1,2}$, Nikolas Makasis$^1$, Monika J. Kreitmair$^1$,\\Zack Xuereb Conti$^{1,2}$, Domenic Di Francesco$^{2,3}$, Ruchi Choudhary$^{1,2}$\\[4pt]
$^1$\textit{\small Energy Efficient Cities Initiative, Cambridge University Engineering Department, Cambridge, UK}\\
$^2$\textit{\small Data-centric Engineering, The Alan Turing Institute, British Library, London, UK}\\
$^3$\textit{\small Computational Statistics \& Machine Learning Group, Cambridge University Engineering Department, Cambridge, UK}\\
\phantom{Line 8}\\
\phantom{Line 9}
}
\date{\vspace{-0.5cm}}	% remove default date and replace the Blank 10th line
\begin{document}

\maketitle

\section*{Abstract}	% Section headings need to be upper and lower case.
\addtocounter{section}{1}
The use of monitored data to improve the accuracy of building energy models and operation of energy systems is ubiquitous, with topics such as building monitoring and Digital Twinning attracting substantial research attention. However, little attention has been paid to quantifying the value of the data collected against its cost. This paper argues that without a principled method for determining the value of data, its collection cannot be prioritised. It demonstrates the use of Value of Information analysis (VoI), which is a Bayesian Decision Analysis framework, to provide such a methodology for quantifying the value of data collection in the context of building energy modelling and analysis. Three energy decision-making examples are presented: ventilation scheduling, heat pump maintenance scheduling, and ground source heat pump design. These examples illustrate the use of VoI to support decision-making on data collection.\\

\section*{Highlights}
\begin{itemize}
\item Value of Information analysis can be used to quantify the value of data collection for improved decision-making.
\item Quantifying data value allows data collection activities to be justified and prioritised.
\item Three example decision problems showcase the range of insights Value of Information analysis can provide on data collection.\\
\end{itemize}

\section*{Keywords}
Value of Information, Bayesian Decision Analysis, Uncertainty quantification, Building energy systems
\newpage
%---------------------------------------------------------------------------------------
\section*{Introduction}
%---------------------------------------------------------------------------------------

% Set the scene of buildings producing large quantities of data
% Data has a cost

Building monitoring systems and the data they gather are used to understand the real behaviour of building energy systems, allowing for substantial improvements to be made in both the operation of the building system, and the design of system retrofits \& new building systems \citep{MOLINASOLANA2017598}. Significant advances in machine learning, artificial intelligence, and compute abilities for design and smart control have driven a rapid increase in the deployment of large, complex, and thereby expensive, sensor networks in building systems \citep{SALIMI2019214}, with many projects ultimately aiming to create a unified Digital Twin which provides information on all aspects of the system. However, such widespread monitoring and data archiving raises significant challenges with regard to the management, handling, security, and meaningful exploitation of the sizeable quantities of data produced. Overcoming these data challenges imposes additional cost to the deployment of already expensive monitoring systems, and raises the question of their utility versus cost.

% We currently don't have a (particularly good) method for rationalising data expenditure (within the building energy field)
% There is therefore potentially/likely some wastage that can be eliminated/there is scope/benefit for having a method to rationalise decisions around information collection for building operation

At present, questions of the economic merit of monitoring data are not commonly asked in the literature. Indeed, there are no studies that rationally quantify the added value of decisions made when designing building monitoring systems. Hence, there may exist significant unidentified wastage in the form of low insight monitoring. This wastage will likely grow if monitoring schemes are not rationalised as building digitisation continues to expand. Further, monitoring investments cannot yet be prioritised in budget constrained systems, as the relative benefit of different monitoring schemes has not been studied in the literature. As a result, the most valuable observations may not be measured, and thus the potential for greater insight and improved decision-making lost.
\newpage

%---------------------------------------------------------------------------------------
\section*{Motivation and Contribution}
%---------------------------------------------------------------------------------------

% VoIA is a mature tool that has been used in many fields
% Brief overview of VoIA literature, applications, different ways it's been used, successes its had
% Highlight the complete lack (justify this with more lit search?) of VoIA applications in the building energy systems literature
% Given the problem highlighted in intro, VoIA has the potential to provide the value add seen in other areas to the building energy community
% Summarise benefits: compare \& prioritise value of different measurements, analyse time value of data (is measurement just for now, or do we want to keep it?)

Assessment of the expected benefit of information gathering through monitoring systems is a critical challenge across a broad range of fields in which decisions must be taken under unknown but observable states, which significantly impact the quality of outcomes, and where the cost of measurement is substantial. Value of Information analysis (VoI) \citep{raiffa1961applied}, a sub-field of Bayesian Decision Analysis, is a mature methodological framework that has been applied extensively to quantify the benefit (or value) of different information gathering schemes in the context of stochastic decision problems. Previous studies have demonstrated that it provides a tractable numerical framework for rationalising information gathering decisions in fields such as medicine, environmental science, engineering, and agriculture \citep{Keisler2013value, Zhang2021value}. VoI has also recently been exploited in the Civil Engineering field to prioritise and justify structural health monitoring, improving the efficiency of asset maintenance operations \citep{Zhang2021value}. The significant benefits of rationalising information gathering tasks achieved in other fields of study motivate the application of VoI to decision problems within the building energy field.

Buildings and their associated energy systems are impacted by a wide range of uncertainties, such as environmental conditions, system characteristics, and user behaviour at a range of time scales. The role of monitoring schemes in buildings is to reduce some of these uncertainties, resulting in improved energy efficiency and user satisfaction.

The contribution of this work is to demonstrate the benefits that the use of VoI could provide in assessing the utility of monitoring schemes in buildings. The paper presents three examples to illustrate both the diversity of decision problems that can be analysed using the VoI framework, and the breadth of insights into the efficacy of monitoring schemes that can be obtained. The three examples span operation, maintenance, and design of energy systems. It is demonstrated that a single, simple VoI metric, the Expected Value of Perfect Information (EVPI), quantifies the merits of a potential measurement, thus enabling the comparison of different measurement schemes and hence their prioritisation, and the assessment of the time value of data (whether data is worth storing, and if so, how much should be kept).\\

\newpage
%---------------------------------------------------------------------------------------
\section*{Value of Information Analysis}
%---------------------------------------------------------------------------------------

\subsection*{Bayesian Decision Analysis}

Bayesian Decision Analysis provides a mathematical framework for studying decision-making in the presence of uncertainties, and seeks to determine the expected optimal action which should be taken by the decision-maker in order to maximise their expected utility.

Consider a generalised stochastic decision problem in which an actor seeks to select a `decision action' to take, $a \in \mathcal{A}$, within an environment with uncertain parameters $\theta$, which have a prior model $\pi(\theta)$. The performance of each available action is given by a utility function which is also dependent upon the uncertain parameters, $u(a,\theta)$. In VoI analysis, before an action $a$ is taken, the actor may choose to take a `measurement action', $e \in E$, from which the actor receives data $z$. The probabilistic model describing the data $f(z|\theta)$ is used to update the prior model, $\pi(\theta)$, to produce a posterior model, $\pi(\theta|z)$, which is then used by the actor to inform their choice of `decision action', and improve their decision making performance.

\begin{figure}[h]
    \centering
    \includegraphics[width=\linewidth]{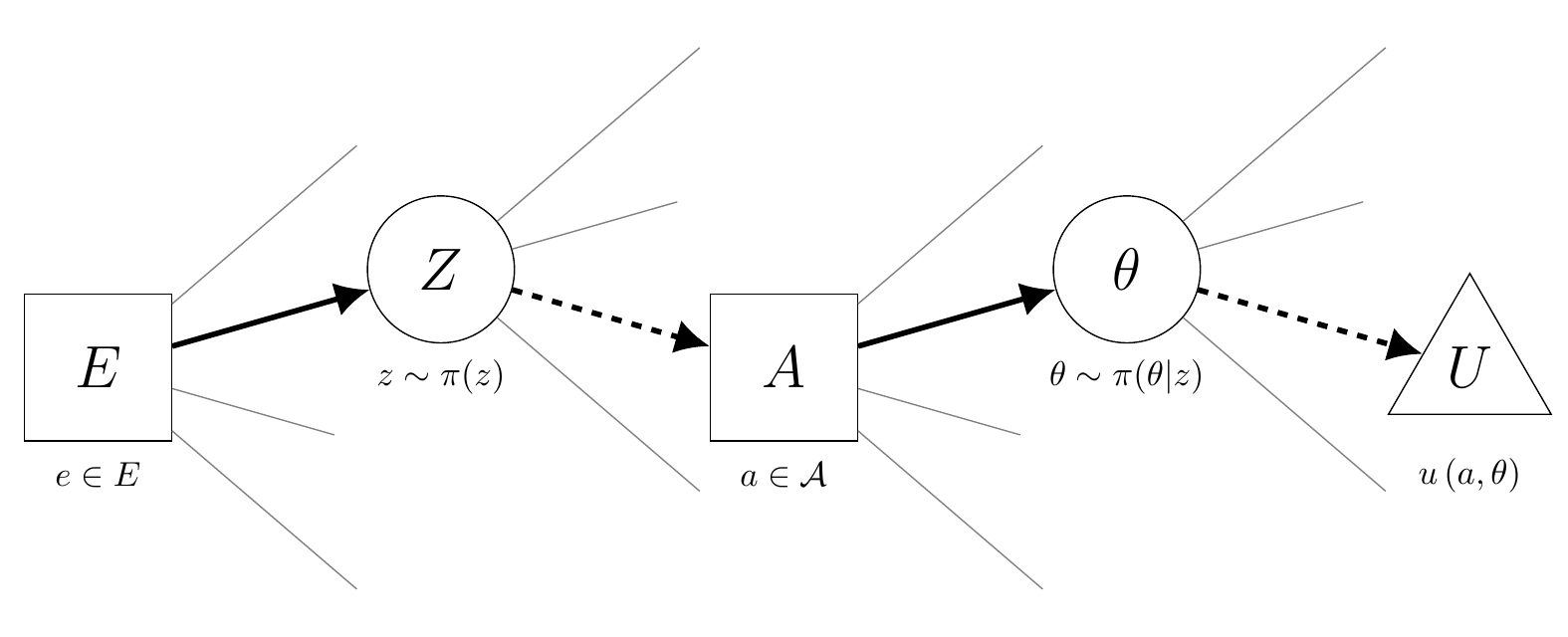}
    \caption{Decision tree representation of Pre-Posterior Decision Problem}
    \label{fig:DT-prepost}
\end{figure}

This generalised model can be represented in decision tree form, as shown in Fig. \ref{fig:DT-prepost}, in which square nodes represent decisions, circular nodes represent uncertainties, and triangular nodes represent utilities.

The actor, who is assumed to be risk neutral, seeks to maximise their expected utility obtained from the selected `decision action'. The actor may choose to do this without taking any measurement. The resulting stochastic optimisation is termed the Prior Decision Problem,
\begin{equation}
    \max_{a \in \mathcal{A}} \, \mathbb{E}_{\theta} \left\lbrace u(a,\theta) \right\rbrace
\end{equation}

and the expected utility achieved by its solution, $a^*$, termed $y^*=\mathbb{E}_{\theta} \left\lbrace u(a^*,\theta) \right\rbrace$.

Alternatively, the decision-maker may additionally consider the selection of a `measurement action' within the decision variables of the optimisation, which leads to the Pre-Posterior Decision Problem,
\begin{equation}
    \max_{e \in E} \, \mathbb{E}_{z} \left\lbrace \max_{a \in \mathcal{A}} \, \mathbb{E}_{\theta|z} \left\lbrace u(a,\theta) \right\rbrace \right\rbrace
\end{equation}

whose solution $a^{**}$ achieves expected utility $y^{**}$.

For the sake of notational simplicity this work will consider the case where only a single non-null `measurement action' is available. However, extensions to multiple measurement actions are straightforward at greater computational expense \citep{DiFrancesco2021decision}.\\

\subsection*{Expected Value of Perfect Information}
The expected Value of Information (VoI) \citep{raiffa1961applied} can be computed using the Bayesian Decision Analysis framework, and is defined as the increase in the expected utility achieved by the optimal action taken by the actor as a result of having additional information at the time of selecting the `decision action'.

In the case where the measurement taken by the actor provides perfect information on the true value of the uncertain parameters of the environment, $\pi(\theta|z) = \delta(\theta-z)$, which allows for the following simplification of the expected utility obtained by the actor making decisions with the help of additional information,
\begin{equation}
\mathbb{E}_{z} \left\lbrace \max_{a \in \mathcal{A}} \, \mathbb{E}_{\theta|z} \left\lbrace u(a,\theta) \right\rbrace \right\rbrace \: \longrightarrow \: \mathbb{E}_{z} \left\lbrace \max_{a \in \mathcal{A}} \, u(a,z) \right\rbrace
\end{equation}

denoted as $y^{**}_p$.\\

The Expected Value of Perfection Information (EVPI) is therefore given by,
\begin{equation}
    \begin{aligned}
        \text{EVPI} &= \mathbb{E}\lbrace \text{perfect information decision utility} \rbrace \\
        & \qquad\quad - \mathbb{E}\lbrace \text{prior decision utility} \rbrace \\
        &= \mathbb{E}_{z} \left\lbrace \max_{a \in \mathcal{A}} \, u(a,z) \right\rbrace - \max_{a \in \mathcal{A}} \mathbb{E}_{\theta} \left\lbrace u(a,\theta) \right\rbrace \\
        &= y^{**}_p - y^*
    \end{aligned}
\end{equation}
%\hfill \\

The EVPI can be interpreted as either the increase in the expected utility the actor can achieve as a result of having perfect knowledge of the uncertain parameters of the environment, or as the reduction in the expected utility arising from the presence of the uncertainties in the environment. Through the former interpretation, the EVPI quantifies the actor's willingness to pay for information that eliminates all uncertainty in the parameter under study from the decision problem. The comparison to the cost of obtaining such a perfect measurement enables the determination of whether such a measurement is economical.\\

\subsection*{Influence Diagrams}

For decision problems comprising larger numbers of nodes with more complex dependency networks, or problems which contain parameters in continuous spaces, influence diagram representations are used instead of decision trees \citep{DiFrancesco2023guidance}. Influence diagrams depict only the causal dependencies between nodes, and not realisation trajectories through the decision graph, and so provide a far clearer description of complex decision problems. In some instances whilst there may exist a decision tree compatible description of the problem, an alternative description containing more nodes and causal dependencies is chosen to provide insight into the physics or dynamics of the system, and so an influence diagram is used. This is the case for the decision problems described below.\\

\subsection*{Extensions}

The VoI framework can be readily extended to further study the impact of uncertainties on decision problems. Such extensions include the study of the value of information from imperfect measurements, which can be shown to be upper-bounded by the EVPI \citep{Keisler2013value}. Also, applications in which multiple measurements schemes are available, the quantification of the expected benefit derived from including stochasticity in the decision optimisation\footnote{Termed the Value of Stochastic Solution (VSS)}, and the quantification and comparison of the impact of different uncertainties on decision-making may be assessed. Further, sensitivity analyses can be performed on VoI calculations to validate the results obtained with respect to the assumptions made in the formulation of the system model.\\

%---------------------------------------------------------------------------------------
\section*{Applications}
%---------------------------------------------------------------------------------------

In this section three decision problems under uncertainty are presented, spanning the operation, management, and design of energy systems. For each decision problem, the EVPI is computed and the insight that the VoI analysis provides on the benefit of measurements is discussed. All VoI computations are performed using 1,000,000 samples from the prior/underlying distribution of the uncertain parameters. The code used to perform the VoI computations for the example problems presented is available at \href{https://github.com/EECi/BSim23-VOI-examples}{\nolinkurl{github.com/EECi/BSim23-VOI-examples}}.\\

\subsection{Building Occupancy Measurement for Real-Time Ventilation Scheduling to Improve Indoor Air Quality}

In mechanically ventilated office spaces, building managers must schedule ventilation system settings to ensure sufficient indoor air quality for the occupants. Adequate ventilation is required to prevent the transmission of airborne infectious diseases such as the SARS-CoV-2 virus (COVID-19), which is both damaging to the health of the occupants and costly for the tenant of the office space through lost productivity. However, operating ventilation at an unnecessarily high rate can impact occupant thermal comfort, lead to excessive carbon emissions, as well as excessive operational cost of the ventilation system through the additional heating/cooling demand required to condition air intake. The risk of viral transmission, and thus the appropriate ventilation setting, is highly dependent on the number of occupants in the office space. However, in the absence of occupant monitoring and dynamic ventilation control systems, building managers must schedule ventilation system settings without knowledge of the exact occupancy level of the space. At the time of deciding ventilation scheduling, the occupancy of the ventilated space over the operation period is uncertain. This uncertainty is particularly relevant in light of recent trends towards `work from home', which reduces the predictability of office occupancy.

This raises the question, ``Would it be worth installing a smart occupancy monitoring and ventilation control system in an office space to improve ventilation scheduling?'', i.e. would the economic benefit of improved ventilation control be greater than the cost of installing such a smart monitoring system?\\

We consider a simple model of indoor air quality in a typical office space. Said office space is taken to have a floor area of 500m$^2$, a ceiling height of 4m, and a maximum occupancy of 55 people. There are four available ventilation settings: 1, 3, 5, and 10 air changes per hour (ACH). The ventilation system is assumed to have a fan of specific power 1.9W/l/s, operating at 60\% efficiency for 10 hours per day. At an electricity unit price of 34 p/kWh the cost of operating the ventilation system at each setting is calculated and provided in Table \ref{tab:vent-costs}.

\begin{table}[b]
\vspace{-5pt}   % Please use appropriate negative vspace to remove the space above/belovw the Table
\caption{Cost of electricity for ventilation system} \label{tab:vent-costs}
\label{tab:tab01}
\centering
\begin{tabular}{| c | c |}
  \hline
  \bf{Vent. Rate (ACH)} & \bf{Energy cost (£/day)}\\
  \hline
  1 & 5.98\\
  3 & 17.94\\
  5 & 29.91\\
  10 & 59.81\\
  \hline
\end{tabular}
\vspace{-5pt}   % Please use appropriate negative vspace to remove the space above/belovw the Table
\end{table}

A model of the probability of viral infection for an individual in an indoor space as described in \citep{de2021evolution} is used, which is also available in web application form at \href{https://airborne.cam/}{\nolinkurl{airborne.cam}} \citep{gkantonas2021airborne}. It is assumed that the base prevalence of infection amongst the occupants is that of the general UK population in February of 2023, 2.18\% \citep{ONS2023infections}, and that infection of an individual leads to 3 days of sick leave, which taking the median daily salary for full-time employees in 2022, costs the tenant £128/day \citep{ONS2022earnings} in lost productivity.

It is further assumed that all occupants are present in the office space for the whole 8 hour work day, that time coupling effects between days can be neglected, i.e. that the model for a single day is representative, and that the prior distribution of occupancy is discrete uniform in the interval 0 to 55 inclusive.

\begin{figure}[h]
    \centering
    \includegraphics[width=\linewidth]{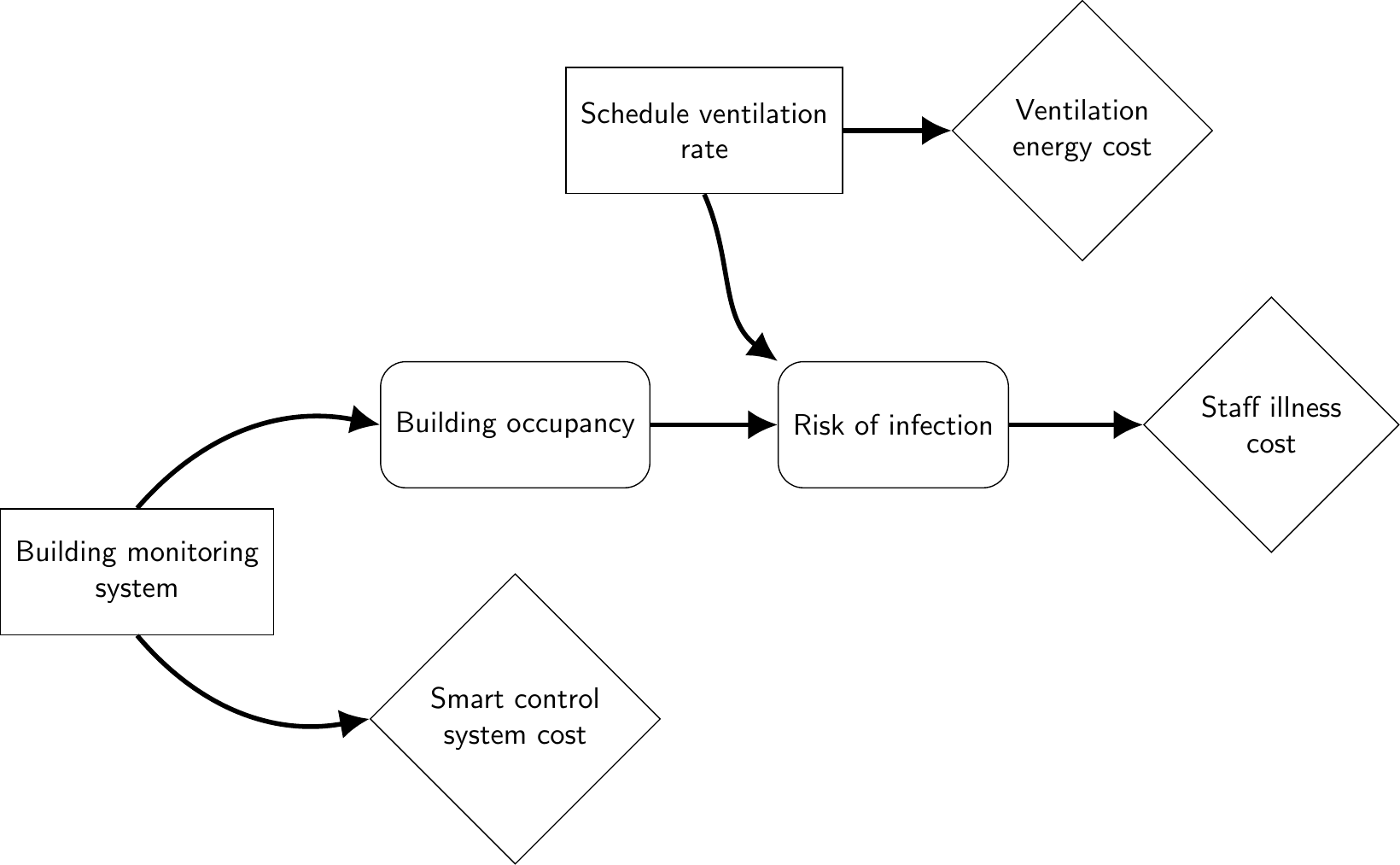}
    \vspace{4pt}
    \caption{Influence diagram representation of building ventilation scheduling decision problem} \label{fig:ID-building-vent}
\end{figure}

The stochastic decision problem is thus to select the ventilation system setting which minimises the sum of the ventilation system operation (electricity) cost, and the cost of lost productivity due to illness to the tenant, subject to uncertainty in the occupancy level of the space. This decision problem can be described via the influence diagram provided in Figure \ref{fig:ID-building-vent}.

Solving the decision problem, the optimal prior decision is found to be a ventilation setting of 5 ACH, with a corresponding expected cost of £72.57/day. However, with a smart monitoring and control system installed, which provides perfect occupancy information, the pre-posterior expected operational cost is £63.15/day. Hence, the EVPI of measuring occupancy level in the context of ventilation scheduling is found to be £9.42/day. Therefore, over a 20 year operational lifetime, the smart monitoring and control system could reduce the target cost by up to £68,800.

The calculated EVPI value can thus aid the decisions on whether investments in monitoring systems are worthwhile. If a proposed smart monitoring and control system is expected to have an installation and maintenance cost over the 20 year lifetime of greater than £68,800, then a building manager can determine that this investment will add net value in the context of ventilation scheduling, and so a cheaper but less precise method of measuring occupancy, such as a desk booking system, may be a more suitable strategy. In this way the EVPI provides a benchmark against which the cost of monitoring systems can be judged to estimate whether they are economically beneficial.

\newpage
\subsection{Degradation Measurement for Optimising Maintenance Scheduling for Air-Source Heat Pumps}

Air-source heat pumps (ASHPs) provide significant advantages to decarbonising the heating of building through their ability to exploit ambient heat in the environment to achieve high Coefficients of Performance (COPs), reducing the direct energy input required to heat a space, and so reducing the embodied carbon emissions of heating. However, through usage, the performance of ASHPs degrade, reducing the COPs they can achieve. Maintenance activities can be undertaken to address this performance degradation and improve the COPs achieved by the heat pumps, reducing the electricity consumed and operational costs. However, regular maintenance is costly.

When deciding how frequently to maintain ASHP units, asset owners seek to trade-off the cost of maintenance activities with the benefits they provide in reduced electricity consumption cost to minimise the total cost of operating the ASHP asset. However, the rate at which the ASHP performance is degrading is typically not known, and so the maintenance scheduling decision must be made under uncertainty in the performance degradation rate of the ASHP.

Smart meter data can be used to better estimate the performance degradation rate of ASHP units, allowing for the scheduling of maintenance to be optimised. But, installing and maintaining smart meters adds additional cost to the operation of the ASHP units. Therefore, asset owners will raise the question, ``Does installing a smart meter on an ASHP unit reduce the overall operating costs by allowing for optimised maintenance scheduling?''.\\

A model of an educational building in the University of Cambridge with 4 identical ASHP units is considered to investigate the economic viability of installing smart meters for maintenance scheduling optimisation. The asset owner selects the number of evenly spaced maintenance activities undertaken per year (the decision variable), $N_m \in \{0,\ldots,12\}$, as to minimise the expected operational cost of the ASHPs.

The annual energy consumed by the ASHP units is given by,
\begin{equation}
    E = \frac{L_H}{\text{SPF}}
\end{equation}
where $L_H$ is the heating load of the building, assumed to be 1.75 GWh/year. The annual Seasonal Performance Factor ($\text{SPF}$) of the ASHPs, the average COP over the heating season, is given by,
\begin{equation}
    \text{SPF} = \text{SPF}'(1-\alpha)(1+\beta)
\end{equation}
$\text{SPF}'$ is the base heat pump SPF, taken to be 3, $\alpha$ is the performance degradation factor (the uncertain parameter for the decision problem), and $\beta$ is given by,
\begin{equation}
    \beta = \frac{\beta_a N_m^\gamma}{\beta_b + N_m^\gamma}
\end{equation}
the parameters $\beta_a$, $\beta_b$, $\gamma$, are empirical parameters with assumed values of 0.05, 2.5, and 1.4 respectively.

The maintenance cost is taken to be £2,210 per activity for all 4 ASHP units \citep{Daikin2022}, and the electricity cost is assumed to be 34 p/kWh.

The degradation parameter $\alpha \geq 0$ is modelled as being distributed as a truncated Normal with mean 0.01 and standard deviation 0.25,
\begin{equation}
    \alpha \sim \mathcal{N}(\alpha, \mu=1e^{-2},\sigma=0.25 : \alpha \geq 0)
\end{equation}

\begin{figure}[h]
    \centering
    \includegraphics[width=\linewidth]{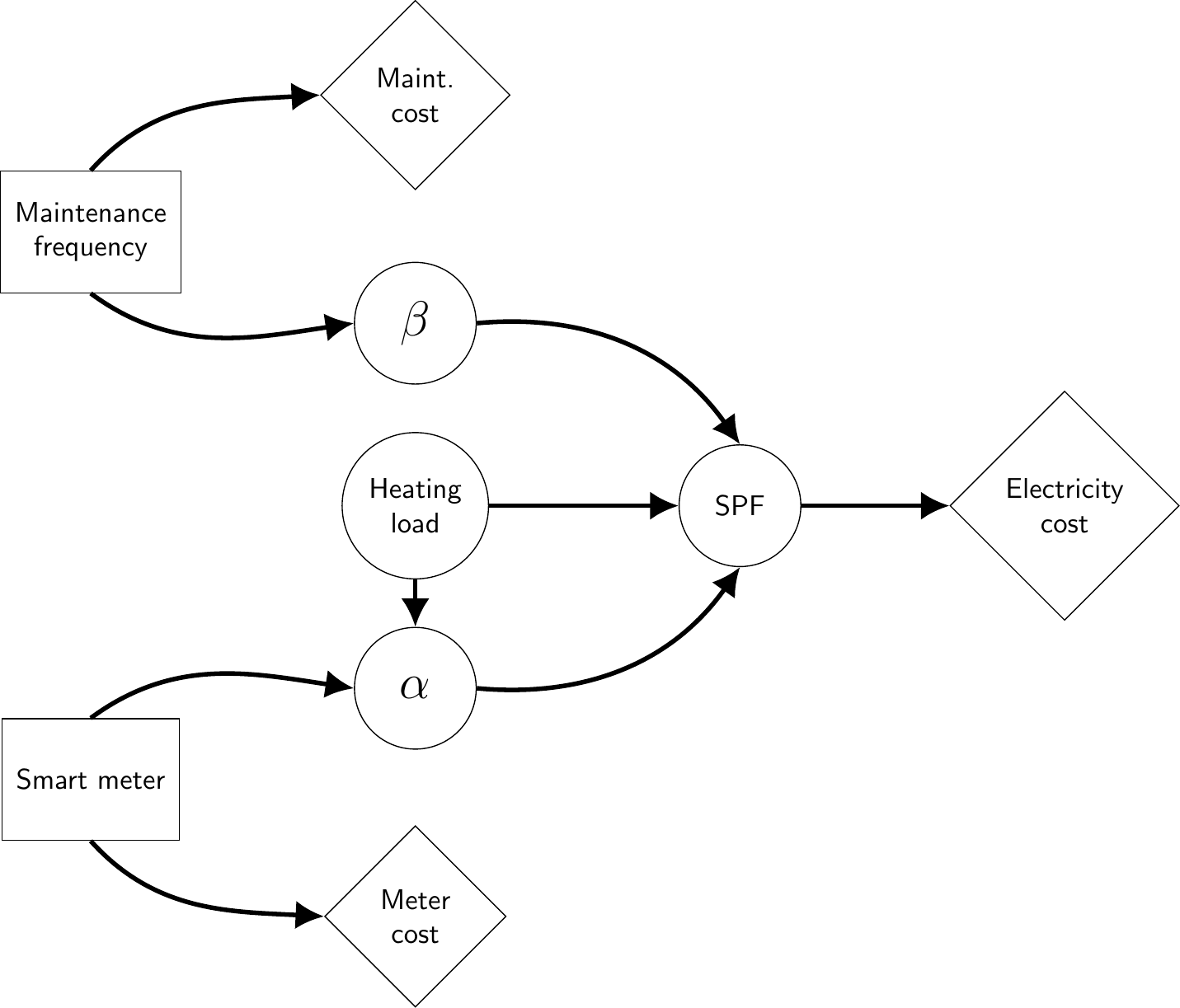}
    \vspace{4pt}
    \caption{Influence diagram representation of ASHP maintenance scheduling decision problem}
    \label{fig:ID-ASHP-maintenance}
\end{figure}

The described stochastic decision problem of optimally scheduling ASHP maintenance is represented in influence diagram form in Figure \ref{fig:ID-ASHP-maintenance}.

Solving the Prior Decision Problem, it is determined that the optimal maintenance frequency is 2 activities per year, leading to an expected cost of £263,120/year. The Pre-Posterior Decision Problem achieves an expected cost of £262,900/year, and so the EVPI of ASHP degradation rate, $\alpha$, is found to be £220/year.

As the cost of operating smart meters monitoring the 4 ASHP units is estimated to be £70/year \citep{Daikin2022}, this analysis demonstrates that smart meters capable of perfectly measuring the performance degradation rate would lead to a net economic benefit of £150/year to the asset owner from the improvements to maintenance scheduling they provide alone. In this way, the asset owner can justify their investment in smart meters. Further, they can estimate the payback time of any capital investments they may have to make through the insight they have gained on the operating cost savings that such a monitoring system could provide.

\subsection{Ground Conductivity Measurement for Optimising Borehole Design in Residential Ground-Source Heat Pump Heat Supply Systems}

Ground-source heat pump (GSHP) systems use the ground as a source and sink of heat to provide cooling and heating for buildings in a highly energy efficient way. As these systems exchange heat with the ground, their performance (characterised by the COPs they achieve) is influenced significantly by the geological properties of the ground with which they exchange heat. In the design of such GSHP heating \& cooling supply systems, it is desired to match the capacity of the GSHP system to energy demand of the building as to minimise the overall cost of operating the supply system over its lifetime. The overall operating cost is composed of the capital cost of constructing the system, and the operational costs, which are primarily the cost of the electricity required to meet the building energy demands. Under-specification of the systems leads to greater electricity usage from less efficient and so higher cost auxiliary heating systems, whilst over-specification of the system results in unnecessary capital cost.

At the time of system design, the thermal properties of the ground are not known precisely, as existing geological survey data provides an uncertain estimate of the ground properties in the site location. However, the system designer has the option to commission a thermal response test at the site location prior to designing the GSHP system. Such tests are time-consuming and incur a significant additional cost, and so ground thermal properties are often incorporated into GSHP design with uncertainty, based on available general information on materials and location. This therefore poses the following question to the designer, ``Would commissioning a thermal response test reduce the overall lifetime cost of the GSHP heat supply system by improving the matching of the designed system capacity to the building load?''.\\

A stylised GSHP system design task for a residential building heat supply system is considered. In this design task, the designer must select the length of boreholes, $L_{\text{bh}}$, to be drilled for the ground heat exchange. The available length choices are 140m to 200m, in 5m increments. The capital cost of borehole drilling is taken to be £70/m/borehole.

It is assumed that the effective ground thermal conductivity, $\lambda_{\text{ground}}$, is the only uncertain geological parameter, and that it is Normally distributed with mean 2W/mK, and standard deviation 0.12W/mK,
\begin{equation}
    \lambda_{\text{ground}} \sim \mathcal{N}(\mu=2,\sigma=0.12)
\end{equation}
which is the average thermal conductivity over the range of borehole depths considered, and covers typical uncertainty ranges given the heterogeneity present in soils.

The designed GSHP system, consisting of 9 boreholes, must supply heat to a small apartment block of 10 flats over a 50 year operational lifetime. The building is assumed to have a typical heating demand distribution for its type, and a load profile is synthesised based on demand values for the UK \citep{MITCHELL2020110240}, and using historic weather data for London. This synthetic load profile, $E_{\text{load}}(t)$, consumes 116MWh/year, or 13.2kW mean load, with a peak load power of 25.2kW.

The model of geothermal borehole operation presented in \citep{LAMARCHE2007188} is used to determine the scheduling of energy extracted from the ground in each time instance, $E_g(t)$, with the borehole fluid temperature, $T_{\text{fluid}}$, constrained to be within the range $5\degree$C to $35\degree$C,
\begin{equation}
    5 \leq T_{\text{fluid}}(t) \leq 35
\end{equation}
In this model, $T_{\text{fluid}}$ is a function of the power extracted from the ground, $E_g(t)$, the ground thermal conductivity, $\lambda_{\text{ground}}$, the borehole length, $L_{\text{bh}}$, and the other assumed ground condition parameters.

Given the fluid temperature schedule determined, the instantaneous COP of the GSHP system is computed using the following empirical relationship from \citep{Kensa2014},
\begin{equation}
    \text{COP}(t) = 4.0279 +0.1319 \cdot T_{\text{fluid}}(t)
\end{equation}
The energy that is provided by the GSHP system to the building is then given by,
\begin{equation}
    E_{\text{GSHP}}(t) = \frac{E_g(t)}{1-1/\text{COP}(t)}
\end{equation}
If the GSHP system is unable to meet the building load in any time instance, the remaining unsatisfied load is provided by an auxiliary heat supply system with a COP of 1.
\begin{equation}
    E_{\text{load}}(t) = E_{\text{GSHP}}(t) + E_{\text{aux}}(t)
\end{equation}
The total electricity consumption of the combined heat supply system is therefore given by,
\begin{equation}
    e_{\text{total}} = \sum_t \left( \frac{E_{\text{GSHP}}(t)}{\text{COP}(t)} + E_{\text{aux}}(t) \right)
\end{equation}
The cost of electricity used by the residential building is taken to be 34 p/kWh.\\

\begin{figure}[h]
    \centering
    \includegraphics[width=\linewidth]{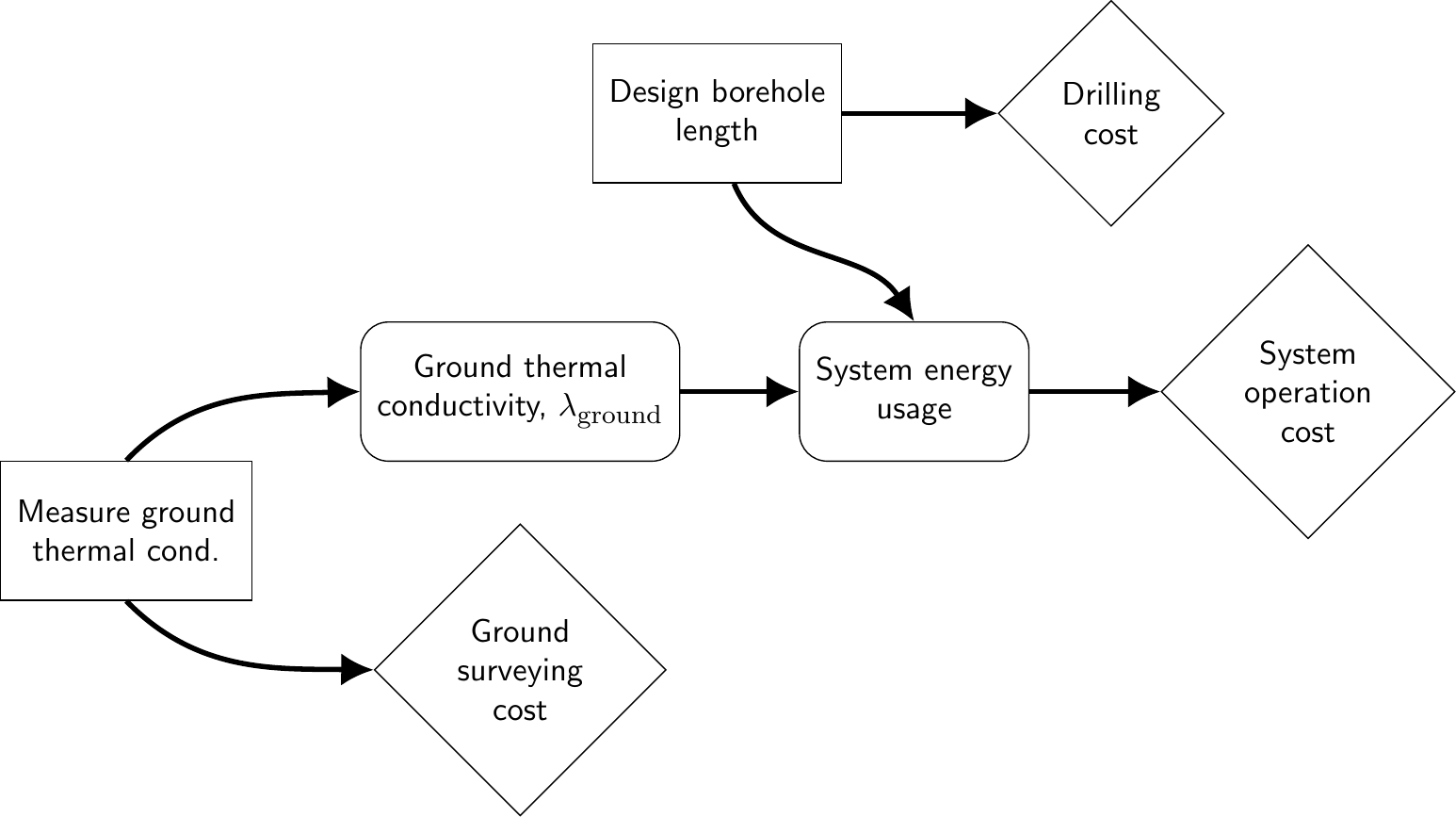}
    \vspace{4pt}
    \caption{Influence diagram representation of GSHP heat supply system design decision problem}
    \label{fig:ID-GSHP-design}
\end{figure}

This stochastic decision problem of designing borehole lengths as to minimise the expected lifetime cost of the heating supply system is represented as an influence diagram in Figure \ref{fig:ID-GSHP-design}.\\
\newpage

The optimal prior decision is found to be a borehole length design of 170m, which results in an expected overall lifetime system cost of £537,400. If a ground condition survey providing a perfect measurement of the ground conductivity were to be conducted before designing the system, then the expected overall cost would be £533,200. Therefore, the EVPI of ground thermal conductivity, $\lambda_{\text{ground}}$, is determined to be £4,200.

The cost of a thermal response test is estimated to be around £5000. Hence, whilst such a survey does improve the system designer's ability to match the capacity of the heat supply system to the building load, this improvement does not warrant the associated cost, and so commissioning one is not an economically sound measurement strategy. This VoI analysis provides the system designer with the insight that the prior distribution information on the ground conductivity already available from historic surveys is sufficient to make an optimal decision, assuming cost risk-neutrality.\\

%---------------------------------------------------------------------------------------
\section*{Discussion}
%---------------------------------------------------------------------------------------
% \begin{itemize}
%     \item Look how many different problems VoIA can tackle and add value to!
%     \item There's a lot more that VoIA can do: imperfect information, sensitivity analysis, etc. (mention extensions targeted in journal paper)
%     \item Discuss caveats of VoIA: computational complexity, problem setup difficulty \& validity, context restrictions \& lower bound on information value
% \end{itemize}

From the example decision problems analysed in the previous section, it can be seen that VoI is a highly flexible framework that can be used to investigate a diverse range of uncertainties affecting building energy systems. Further, the broad scope for interpretation of the VoI metric values computed allows for many different aspects of the impact of uncertainties on decision making in building energy systems to be investigated, and so many different insights into the role of data collection in enhancing decision making to be gained. These insights enable improved decision making for the design of monitoring systems.

However, care must be taken in the interpretation of VoI metric values to ensure that the proposed insights are valid and can be justified under the assumptions of the VoI framework. Due to its definition, the EVPI quantifies the expected utility benefit achieved by eliminating an uncertainty from the decision problem, by providing perfect information on its value. But, as no physical measurements can be perfect, this therefore provides an upper-bound on the value of measuring a given quantity before undertaking a decision. A limitation of the EVPI calculation is that it cannot determine the value of a real measurement of the quantity, but nonetheless when interpreted correctly, useful insight can be gained. Further, whilst the value of a given data collection for a set of decisions can be computed, a mathematical formulation of each decision is required. Hence, the benefits arising from unforeseen or unmodelable future decisions will not be considered in the VoI calculation. % Further, as VoI computes the value of information acquisition in the context of a particular decision problem, a single calculation cannot determine the overall value of that measurement or information, as said data may provide additional value to other decision problems within the same building energy system, including unforeseen future problems.
Finally, as VoI does not require any measurements to be undertaken, the value determine by a VoI analysis is only valid within the context of the system model used to compute it. Therefore, if that model provides a poor representation of the physical system it attempts to model, then the insights gained from the VoI analysis may not map onto the true building energy system, and erroneous decision strategies may result. Hence, accurate VoI estimates can only be achieved in systems for which a reliable model (either mathematical or simulation based) exists.

The extensions of the VoI framework discussed in the theory section allow for more precise insights into the impact of uncertainties on building energy related decision problems to be achieved, and in doing so enable the limitations of the VoI calculations discussed above to be addressed. Through the use of imperfect information analyses, the value of real, imprecise, measurements of physical quantities can be determined, as well as the trade-off between measurement precision (and its associated cost) and the expected improvements in decision making performance achieved. % If a given measurement is able to inform multiple decision problems, then coupled VoI analyses can be performed to determine the overall value of said information to the set of problems for which it is used. And for ...
For building systems where there is uncertainty in the model assumptions, either these uncertainties can be introduced into the VoI analysis performed, or sensitivity analysis can be applied to the VoI calculations to determine the assumption conditions under which the proposed insights hold. Further work is required to understand the insights that can be achieved using these more complex VoI metrics in the context of building energy systems.

The key challenge of VoI is formulating a decision problem that both represents the physical building energy system under study sufficiently accurately, and yields VoI metric values whose interpretation provides meaningful insight into the impact of the uncertainties in the system on the decision problem.
Whilst the statistical analysis of the VoI framework provides valuable insights into the impact of uncertainties, these come at a significant computational expense, as the decision model must be evaluated many times to provide a Monte Carlo approximation with reasonable error. However, the development of methods to improve the computational efficiency of VoI calculations is an active area of research.

%---------------------------------------------------------------------------------------
\section*{Conclusion}
%---------------------------------------------------------------------------------------
% VoIA has potential for significant value-add by can providing a method for rationalising decisions around information collection in building energy problems, which can be used to tackle the current deluge of data being/planned to be collected and justify the associated cost.
% Even the most simple VoI metric, EVPI, can provide breadth of insights.
% This is a hammer which would be very useful to this community as we have identified all these nails, and we think there are lots more.

Value of Information analysis provides a justifiable and auditable framework for addressing questions of the economic benefit of data collection in building energy systems, and could enable the rationalisation of data collection strategies, reducing resource wastage on low insight data. This rationalisation of data collection is particularly pertinent in light of the rapidly growing deployment of smart monitoring systems in new buildings and retrofits, and the volumes of data they could produce. This work has demonstrated the broad scope of problems to which VoI can be applied across the building energy systems field, the flexibility of the framework and the diversity of insights that it can provide to support data collection decision making, and the economic benefits that can be achieved through its application. It is proposed that significant further research effort is required to identify decision problems within energy systems design and management to which VoI could be applied, and exploit the insights it provides to improve the efficiency of data usage within the field.\\

%---------------------------------------------------------------------------------------
% References
\bibliographystyle{bs2023}
\bibliography{main}
%---------------------------------------------------------------------------------------

\end{document}